\documentclass[conference]{IEEEtran}
\IEEEoverridecommandlockouts
\usepackage{cite}
\usepackage{amsmath,amssymb,amsfonts}
\usepackage{algorithmic}
\usepackage{graphicx}
\usepackage{textcomp}
\usepackage{xcolor}
\usepackage{mathrsfs}
\usepackage{cases}
\usepackage{multirow}
\usepackage{etoolbox}

\def\BibTeX{{\rm B\kern-.05em{\sc i\kern-.025em b}\kern-.08em
    T\kern-.1667em\lower.7ex\hbox{E}\kern-.125emX}}

\makeatletter
\newcommand{\linebreakand}{%
  \end{@IEEEauthorhalign}
  \hfill\mbox{}\par
  \mbox{}\hfill\begin{@IEEEauthorhalign}
}
\makeatother

\makeatletter
\patchcmd{\@maketitle}
  {\addvspace{0.5\baselineskip}\egroup}
  {\addvspace{-0.5\baselineskip}\egroup}
  {}
  {}
\makeatother

\begin{document}

\title{Dynamic Microgrid Formation Considering Time-dependent Contingency: A Distributionally Robust Approach
\\
\thanks{The work was supported by the Shenzhen Science and Technology Innovation Commission through the project ``Research on Optimization Theory and Strategies for Load Restoration of Resilient Distribution Networks'' (JCYJ20220530143010024).}
}
\author{\IEEEauthorblockN{
Ziang Liu\textsuperscript{1}, 
Sheng Cai\textsuperscript{2}, 
Qiuwei Wu\textsuperscript{1}, 
Xinwei Shen\textsuperscript{1}, 
Xuan Zhang\textsuperscript{1},
Nikos Hatziargyriou\textsuperscript{3}}
\IEEEauthorblockA{
\textsuperscript{1}\textit{Tsinghua Shenzhen International Graduate School,} \textit{Tsinghua University, Shenzhen, China} \\
\textsuperscript{2}\textit{School of Automation, Nanjing University of Science and Technology, Nanjing, China} \\
\textsuperscript{3}\textit{School of Electrical \& Computer Engineering, National Technical University of Athens, Athens, Greece} \\
liuziang1119@163.com} \\
}
\maketitle
\begin{abstract}
The increasing frequency of extreme weather events has posed significant risks to the operation of power grids. 
During long-duration extreme weather events, microgrid formation (MF) is an essential solution to enhance the resilience of the distribution systems by proactively partitioning the distribution system into several microgrids to mitigate the impact of contingencies. This paper proposes a distributionally robust dynamic microgrid formation (DR-DMF) approach to fully consider the temporal characteristics of line failure probability during long-duration extreme weather events like typhoons. 
The boundaries of each microgrid are dynamically adjusted to enhance the resilience of the system.
Furthermore, the expected load shedding is minimized by a distributionally robust optimization model considering the uncertainty of line failure probability regarding the worst-case distribution of contingencies. The effectiveness of the proposed model is verified by numerical simulations on a modified IEEE 37-node system.
\end{abstract}

\section{Introduction}
Nowadays, climate changes are increasing the frequency and intensity of extreme weather events and natural disasters. Power systems have been seriously affected by severe weather events over the past few years. For example, an unprecedented snowstorm struck Texas in 2021, leading to a large area power outage in a few days affecting millions of people\cite{Texas_Snowstorm}. These extreme events heavily impact the distribution system for its vulnerability. Statistical data indicates that distribution system failures are responsible for 90\% of customer outage minutes in the United States\cite{campbell2012weather}. 

The integration of distributed energy resources (DERs) and intelligent devices has transformed conventional passive distribution systems into active distribution systems, which enables the distribution system operators to proactively schedule the system to enhance resilience. Existing studies indicate that proactive allocations of DERs and remote controlled switches (RCSs) can significantly mitigate load shedding during extreme weather events\cite{lei2017remote}. 
Also, the network reconfiguration method can control tie-line switches to form a new network topology to supply critical loads after contingencies to enhance resilience\cite{shi2021network}.
Furthermore, microgrids with various types of distributed generators (DGs) have the potential to ensure the power supply of critical loads and facilitate system resilience under natural. disasters\cite{ding2017resilient,hussain2019microgrids}.

Microgrid formation (MF) is one critical procedure in resilience enhancement of the system which defines the boundaries of each microgrid in the system. MF can be classified into static microgrid formation (SMF) and dynamic microgrid formation (DMF). For SMF, the distribution system is partitioned into several microgrids with static boundaries indicating that the topology of each microgrid is fixed\cite{wang2015networked,hussain2016resilient}. For DMF, the boundaries of each microgrid are determined dynamically which can provide more operational flexibility for the system\cite{kim2016framework,lei2020radiality,cai2022hybrid}. Furthermore, DMF can better address the self-sufficiency of MGs in the presence of line contingencies and time-varying power demand\cite{cai2022hybrid}.

To deal with the uncertainty of the contingencies, robust optimization (RO) and stochastic programming (SP) are two common approaches to solving the problem. Specifically, RO-based models find the worst-case scenario based on the \textit{N-k} security criterion to give a conservative solution, which gives up the accessible distribution information of line failure probability. The RO approach may yield an overly conservative solution that restricts the power supply of critical loads during extreme events\cite{gholami2017proactive}. The SP-based models optimistically adopt the failure probabilities of each contingency, and generate a set of scenarios representing the uncertainty of contingencies to make decisions. However, the large number of scenarios can dramatically increase the computational complexity. Also, the failure probability of each line is not accurate which may lead to overly optimistic solutions. Thus, both RO and SP suffer limitations from their very modeling premises.

Distributionally robust optimization (DRO) is an important generalization of RO and SP, which provides a powerful modeling framework and resolves the limitations of both RO and SP. It has been applied to several resilience enhancement problems like network reconfiguration, and microgrid formation. 
In \cite{babaei2020distributionally}, the DRO model optimally reconfigures the network topology considering the contingencies with low probability and high impacts.
In \cite{cai2021distributionally}, DRO is applied to the post-disaster microgrid formation to maximize the load restoration concerning the worst distribution of subsequent contingencies. However, the temporal dynamic impact of the extreme weather events, like the path of a typhoon, is neglected which may result in improper microgrid formation with low efficiency. To the best of our knowledge, the deployment of DRO in DMF considering the uncertainty of contingencies is rarely conducted in previous research.

In this paper, a distributionally robust dynamic microgrid formation (DR-DMF) method is proposed, in which the temporal characteristics of line failure probability are fully considered during long-duration extreme weather events like typhoons. 
The resilience of the distribution system is enhanced by proactively partitioning the system into several microgrids, and adjusting the boundaries of microgrids dynamically.
Furthermore, the expected load shedding is minimized by a distributionally robust optimization approach considering the uncertainty of line failure probability regarding the worst distribution of contingencies in each time interval. Finally, the proposed model is verified on a modified IEEE 37-bus system with tie-line switches.
\vspace{-0.3em}
\section{Problem Statement}
\vspace{-0.2em}
\begin{figure}[b]
\vspace{-0.5em}
    \centering
    \includegraphics[width=0.9\linewidth]{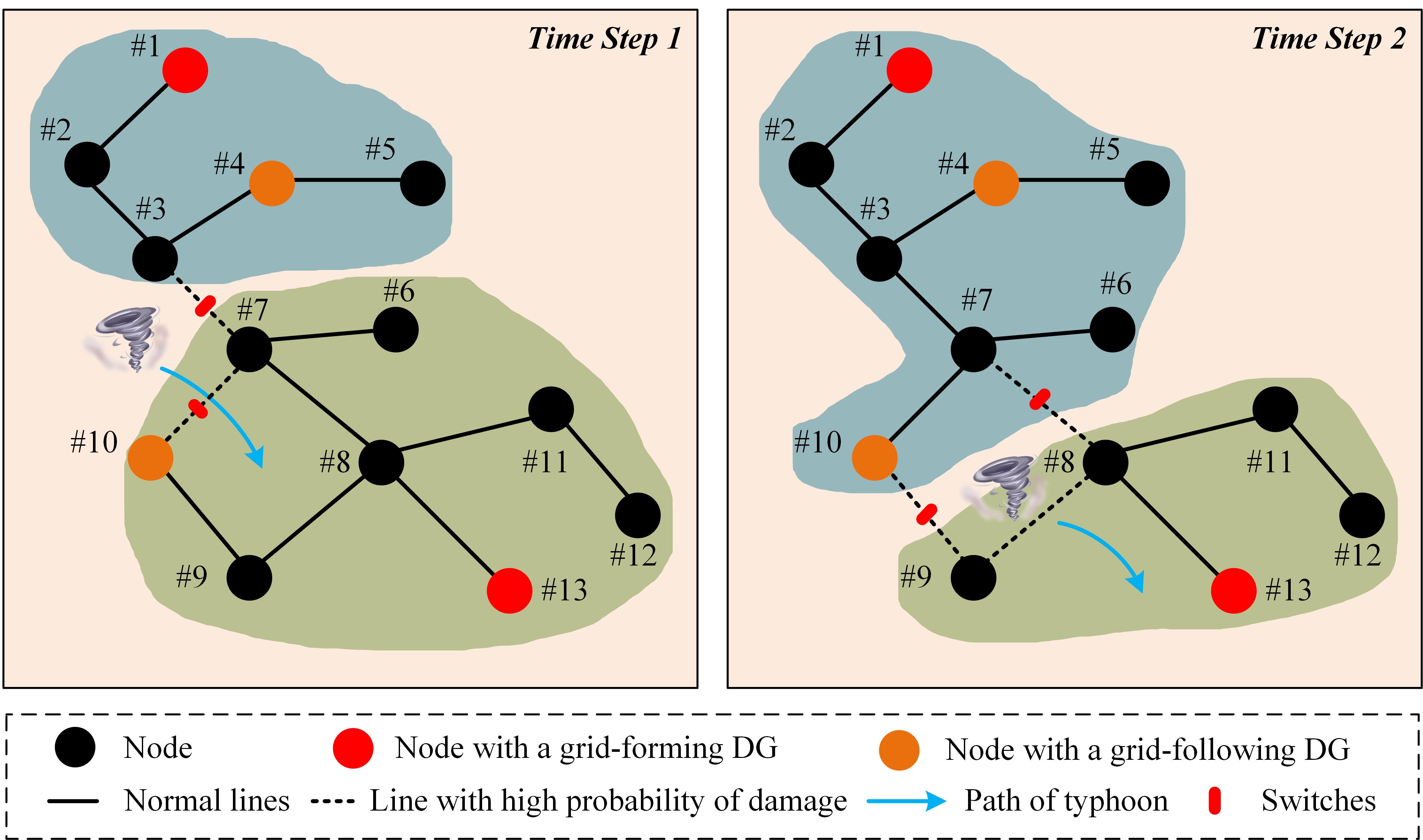}
    \caption{An illustrative example for dynamic microgrid formation}
    \label{fig_illu_example}
\end{figure}
\begin{figure}[t]
    \centering
    \includegraphics[width=0.9\linewidth]{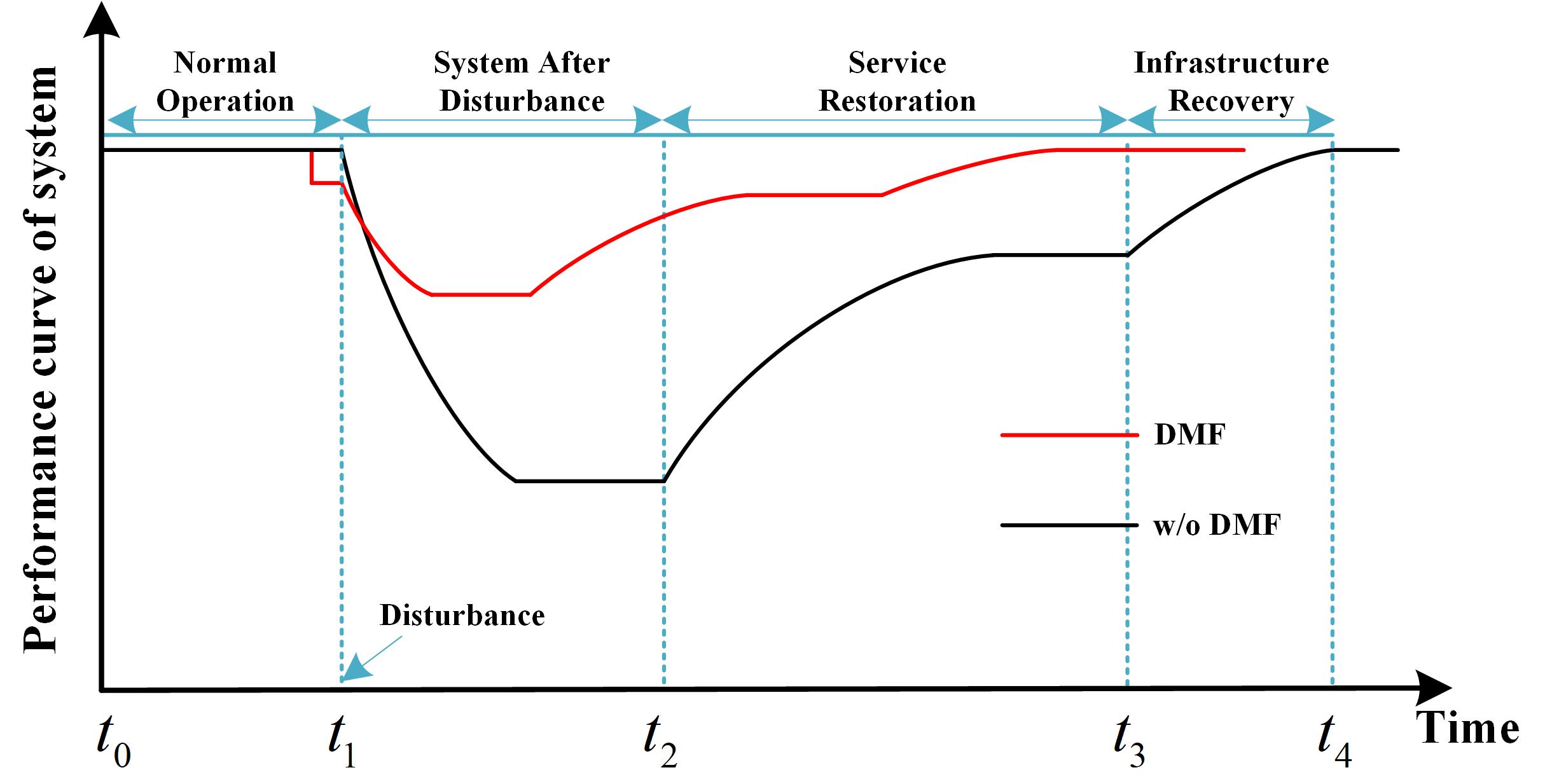}
    \caption{Typical response of a resilient system after a disruption}
    \label{fig_resilience_curve}
    \vspace{-1em}
\end{figure}
Fig. \ref{fig_illu_example} is an example to illustrate the dynamic microgrid formation. It is assumed that the distribution system is disconnected from the upstream grid. The path of the typhoon is assumed to be predictable, and the inaccurate line failure probability is also known. 
Before the fault occurs, the system is partitioned into small microgrids, and the boundaries of each microgrid are dynamically formed based on the result of DR-DMF to mitigate the impact of the probable subsequent contingencies during the long-duration extreme weather events. 
Power system resilience describes the performance of the system in extreme scenarios, including the worst-case scenario. The process of a resilient power system through disruption can be illustrated with the resilience triangle which is shown in Fig. \ref{fig_resilience_curve}. The proactive dynamic microgrid formation can mitigate the impact of subsequent contingencies by avoiding chain failures and a large amount of load shedding, and the performance is shown as the red curve in Fig. \ref{fig_resilience_curve}.

\section{Model Formulation}
\subsection{Objective function}
The prime objective of the DR-DMF model is to minimize the total load shedding according to its importance and ensure the power supply of critical loads. 
To consider the subsequent contingencies, the ambiguity set of line status $\mathscr{P}$ is constructed.
The objective function of DR-DMF is shown in (\ref{eq_objective}).
\begin{gather}
    \min_{\mathbf{x} \in \mathbf{X}} \max_{\mathbb{P} \in \mathscr{P}} \mathbb{E}_{\mathbb{P}} [Q(\mathbf{x},\mathbf{u})]
    \label{eq_objective}
    \\
    Q(\mathbf{x},\mathbf{u})=\min_{\mathbf{y\in \mathbf{G}(\mathbf{x},\mathbf{u})}} \sum_{t\in \mathcal{T}}\sum_{i\in \mathcal{N}} w_i PL_{i,t}
    \label{eq_obj_Q}
\end{gather}
where the objective function decides the microgrid formation ($\mathbf{x}$:\textit{binary variables}) in the first stage, and seeks the worst distribution of contingencies ($\mathbf{u} \sim \mathbb{P}$) in the second stage. $\mathbb{E}_{\mathbb{P}} [Q(\mathbf{x},\mathbf{u})]$ is the expected value of weighted load shedding by optimizing the operation process ($\mathbf{y}$:\textit{continuous variables}), the detailed formulation of $Q(\mathbf{x},\mathbf{u})$ is shown in (\ref{eq_obj_Q}), where $w_i$ and $PL_{i,t}$ represent the weighted coefficient and the power load. 

\subsection{Constraints}
The DR-DMF model is subjected to two types of constraints: microgrid topology constraints ($\mathbf{X}$) and post-events scheduling constraints ($\mathbf{G}$), corresponding to first-stage constraints and second-stage constraints, respectively.
\subsubsection{Microgrid topology constraints}
The radiality of each microgrid is guaranteed by the combination of fictitious flow model\cite{lavorato2011imposing} and spanning tree model\cite{jabr2012minimum}. 
\begin{gather}
    \begin{aligned}
        \sum_{j|(i,j)\in \mathcal{E}} f_{ij,t} - \sum_{j|(i,j)\in \mathcal{E}} f_{ji,t} \leq 0, \quad \forall i \in \mathcal{N}_m, t\in \mathcal{T}
        \label{eq_fictitious_flow_source}
    \end{aligned}
    \\
    \sum_{j|(i,j)\in \mathcal{E}} f_{ij,t} +\hspace{-0.5em} \sum_{j|(i,j)\in \mathcal{E}} f_{ji,t} + sv^{tp}_{i,t} = 1, \forall i \notin \mathcal{N}_m, t\in \mathcal{T}
    \label{eq_fictitious_flow_balance}
    \\
    0 \leq f_{ij,t} \leq c_{ij,t} \cdot M, \quad \forall (i,j)\in \mathcal{E}, t\in \mathcal{T}
    \label{eq_fictitious_y}
    \\
    c_{ij,t}=0, \quad \forall i\in \mathcal{G}^b, (i,j)\in \mathcal{E}, t\in \mathcal{T}
    \label{eq_st1}
    \\
    c_{ij,t} + c_{ji,t} \leq 1, \quad \forall (i,j)\in \mathcal{E}, t\in \mathcal{T}
    \\
    \sum_{j|(i,j)\in \mathcal{E}} c_{ji,t} + sv_i^{tp}=1, \quad \forall i \notin \mathcal{N}_m, t\in \mathcal{T}
    \label{eq_st3}
    \\
    0 \leq sv_{i,t}^{tp} \leq 1, \quad \forall i\in \mathcal{N}, t\in \mathcal{T}
    \label{eq_slack_var}
\end{gather}
\begin{gather}
    c_{ij,t}+c_{ji,t} = c_{ij,t-1}+c_{ji,t-1} + v_{ij,t}^{cl} - v_{ij,t}^{op} \quad \forall (i,j)\in \mathcal{E}
    \label{eq_switch}
    \\
    \sum_{(i,j)\in \mathcal{E}} v_{ij,t}^{cl} + v_{ij,t}^{op} \leq N^{sw,max},\quad \forall t\geq 2
    \label{eq_switch_times}
\end{gather}
where $f_{ij,t}$ is the fictitious flow of line $ij$, $c_{ij,t}$ stands for the parent-child relationship between node $i$ and $j$, $sv^{tp}_{i,t}$ is the slack variable to ensure the feasibility of the model. 
To fully harness the ability of grid-forming DGs, each grid-forming DG controls a single microgrid, indicating that the number of microgrids is equal to the number of grid-forming DGs in the distribution system.
Fictitious flow can only be supplied by the grid-forming generators of each microgrid which is shown in (\ref{eq_fictitious_flow_source}). Fictitious flow balance is guaranteed by (\ref{eq_fictitious_flow_balance}). Equation (\ref{eq_fictitious_y}) indicates that the fictitious flow can only be transmitted from the parent node to the child node. 
Equations (\ref{eq_st1})-(\ref{eq_st3}) are the spanning tree model. 
Equation (\ref{eq_slack_var}) specifies that the slack for the balance of fictitious flow at node $i$ is confined to node $i$. It ensures that $sv_{i,t}^{tp}$ stands for the connection status of node $i$ at time step $t$.
Equation (\ref{eq_switch}) indicates the switch action of each line between two time steps.
Furthermore, the switch of the lines can not be frequently changed, so the numbers of switch actions need to be limited, which is shown in (\ref{eq_switch_times}).
Slack variables are introduced to ensure the feasibility under extreme conditions in (\ref{eq_fictitious_flow_balance}) and (\ref{eq_st3}).

\subsubsection{Post-events operation constraints}
The system operation constraints need to be constructed for each extreme scenario $\mathbf{u}$ in the second stage. The second stage optimization can be solved with these constraints to get the value of $Q(\mathbf{x^*},\mathbf{u^*})$ under microgrid formation decision $\mathbf{x^*}$ and scenario $\mathbf{u^*}$.
\begin{gather}
    \begin{aligned}
        PG_{i,t} \leq (1-sv_{i,t}^{tp}) \cdot PG_{i,t}^{max} , \quad \forall i\in \mathcal{G}, t\in \mathcal{T}
        \\
        QG_{i,t} \leq (1-sv_{i,t}^{tp}) \cdot QG_{i,t}^{max} , \quad \forall i\in \mathcal{G}, t\in \mathcal{T}
        \label{eq_DG_output}
    \end{aligned}
    \\
    \begin{aligned}
        \sum_{j|(i,j)\in \mathcal{E}} PF_{ij,t} + PL_{i,t} = \sum_{j|(i,j)\in \mathcal{E}} PF_{ji,t} + PG_{i,t}, \\ 
        \sum_{j|(i,j)\in \mathcal{E}} QF_{ij,t} + QL_{i,t} = \sum_{j|(i,j)\in \mathcal{E}} QF_{ji,t} + QG_{i,t}, \\
        \quad \forall i \in \mathcal{N}, t\in \mathcal{T}
        \label{eq_powerflow_start}
    \end{aligned}
    \\
    \begin{aligned}
        V_{i,t}=V_{j,t}+(R_{ij}PF_{ij,t}+X_{ij}QF_{ij,t})+\delta_{ij,t},
        \\
        \forall (i,j)\in \mathcal{E}, t\in \mathcal{T}
    \end{aligned}
    \\
    \begin{aligned}
            -1+(c_{ij,t}+c_{ji,t}) \leq \delta_{ij,t} \leq 1-(c_{ij,t}+c_{ji,t}), 
            \\ \forall (i,j)\in \mathcal{E}, t\in \mathcal{T}
            \label{eq_powerflow_end}
    \end{aligned}
    \\
    V_{i,t}=1, \quad \forall i\in \mathcal{N}_m, t\in \mathcal{T}
    \label{eq_slackbus}
    \\
    \begin{aligned}
            (1-sv_{i,t}^{tp}) \cdot V^{min} \leq V_{i,t} \leq (1-sv_{i,t}^{tp}) \cdot V^{max}, \\ \forall i\in \mathcal{N}, t\in \mathcal{T} 
            \label{eq_vol_lim}
    \end{aligned} 
    \\
    \begin{aligned}
        -(c_{ij,t}+c_{ji,t})\cdot M \leq PF_{ij,t} \leq (c_{ij,t}+c_{ji,t}) \cdot M, 
        \\
        \forall (i,j)\in \mathcal{E}, t\in \mathcal{T}
        \label{eq_powerline_lim_1}
    \end{aligned}
    \\
    \begin{aligned}
        -u_{ij,t} \cdot M \leq PF_{ij,t} \leq u_{ij,t} &\cdot M, \\ 
        &\forall (i,j)\in \mathcal{E}, t\in \mathcal{T}
        \label{eq_powerline_lim_2}
    \end{aligned}
\end{gather}
where $PG_{i,t}$, $QG_{i,t}$, $PL_{i,t}$, $QL_{i,t}$, $PF_{ij,t}$, $QF_{ij,t}$, and $V_{i,t}$ represent the active and reactive power of generation, load, power flow, and voltage, respectively, $\delta_{ij,t}$ is an auxiliary variable that describes the voltage relation between two end-nodes of a line. The active and reactive power output of DGs is shown in (\ref{eq_DG_output}), indicating that only the DGs belonging to a microgrid can provide power. For the power flow of the system, the LinDistFlow model is employed to mitigate the computational burden of the model, which is shown in (\ref{eq_powerflow_start})-(\ref{eq_powerflow_end}).
Since different microgrids are isolated, it is reasonable to treat each grid-forming DG bus as a slack bus, which is shown in (\ref{eq_slackbus}). Equation (\ref{eq_vol_lim}) indicates that only the nodes belonging to a microgrid should satisfy the voltage limit. Equation (\ref{eq_powerline_lim_1}) and (\ref{eq_powerline_lim_2}) ensure that power transmission is only possible when the line is closed and remains undisturbed simultaneously.

\section{Solution Methodology}

In this section, the reformulation and solution methodology for the DR-DMF model are introduced. To simplify the expression, we rewrite the model in the compact form as below:
\begin{subequations}
    \begin{align}
        \min_{\mathbf{x}} &\max_{\mathbb{P} \in \mathscr{P}} \mathbb{E}_{\mathbb{P}} [Q(\mathbf{x},\mathbf{u})]
        \label{eq_obj_compact}
        \\
        s.t. &\quad \mathbf{D}\mathbf{x} \leq \mathbf{h} \quad (\mathbf{X})
        \\
        &\quad \mathbf{F}\mathbf{y}  \leq \mathbf{b}-\mathbf{E}\mathbf{x} - \mathbf{H}\mathbf{u} \quad (\mathbf{G(x,u)})
    \end{align}
    \label{eq_compact}
\end{subequations}
The ambiguity set $\mathscr{P}$ is constructed with the \textit{N-k} security criterion and the failure probability of lines, as shown below:
\begin{gather}
    \mathscr{P} := \{\mathbb{P}\in \mathcal{D} :0\leq \mathbb{E}[1-u_{ij,t}]\leq \mu^{max}_{ij}\}
    \\
    \mathcal{D} := \{\sum_{(i,j)\in \mathcal{E}} (1-u_{ij,t}) \leq k, u_{ij,t}\leq u_{ij,t-1}\} 
\end{gather}
Then, we can rewrite the inner-level of (\ref{eq_obj_compact}) as follows:
\begin{subequations}
    \begin{align}
        \max_{\mathbb{P}} &\int_\mathcal{D} Q(\mathbf{x},\mathbf{u}) d\mathbb{P}
        \\
        s.t. &\int_\mathcal{D} d\mathbb{P} = 1
        \label{eq_int_con1}
        \\
        &\int_\mathcal{D} (1-u_{ij,t}) d\mathbb{P} \leq \mu^{max}_{ij}
        \label{eq_int_con2}
    \end{align}
\end{subequations}
Since there exists a $\mathbb{P}$ that satisfies (\ref{eq_int_con1}) and (\ref{eq_int_con2}), the \textit{Slater condition} holds, indicating that the strong duality is satisfied. Thus, we can reformulate (\ref{eq_obj_compact}) into its dual form:
\begin{gather}
    \min_{\mathbf{x} \in \mathbf{X},\mathbf{\alpha}, \mathbf{\beta}\geq 0} \mathbf{\alpha} + \sum_{t\in \mathcal{T}} \sum_{(i,j)\in \mathcal{E}} \mu_{ij}^{max}\beta_{ij,t}
    \\
    s.t. \quad \mathbf{\alpha} + \sum_{t\in \mathcal{T}} \sum_{(i,j)\in \mathcal{E}} (1-u_{ij,t})\beta_{ij,t} \geq Q(\mathbf{x},\mathbf{u}),\quad \forall \mathbf{u}\in \mathcal{D} 
\end{gather}
where $\mathbf{\alpha}$ and $\mathbf{\beta}$ are the dual variables associated to (\ref{eq_int_con1}) and (\ref{eq_int_con2}). We can further reformulate the problem by eliminating variable $\mathbf{\alpha}$, shown as follows:
\begin{gather}
    \min_{\mathbf{x} \in \mathbf{X},\mathbf{\beta}\geq 0} \max_{\mathbf{u}\in \mathcal{D}} \{Q(\mathbf{x},\mathbf{u})+\sum_{t\in \mathcal{T}} \sum_{(i,j)\in \mathcal{E}} (\mu_{ij}^{max}-1+u_{ij,t})\beta_{ij,t}\}
\end{gather}
Finally, substituting $Q(\mathbf{x},\mathbf{u})$ with (\ref{eq_obj_Q}), we can get the final formulation of the DR-DMF model which is shown as follows:
\begin{gather}
    \begin{aligned}
        \min_{\mathbf{x}\in \mathbf{X},\mathbf{\beta}\geq 0} (\mu_{ij}^{max}-1)\beta_{ij,t} + \max_{\mathbf{u}\in \mathcal{D}} \min_{\mathbf{y\in \mathbf{G}(\mathbf{x},\mathbf{u})}}\{\sum_{t\in \mathcal{T}}\sum_{i\in \mathcal{N}} w_i L_{i,t}
        \\
        +\sum_{t\in \mathcal{T}} \sum_{(i,j)\in \mathcal{E}} u_{ij,t}\beta_{ij,t}\}
    \end{aligned}
    \label{eq_final_obj}
\end{gather}

The reformulation of the DR-DMF model becomes a conventional two-stage robust model, which can be effectively solved using well-known approaches like the column-and-constraint generation (C\&CG) method. The convergence efficiency has been proven in \cite{zeng2013solving}. The dual form of the inner-level contains bilinear terms, and it can be linearized by the McCormick method introduced in \cite{mccormick1976computability}.
\begin{figure}[b]
    \centering
    \includegraphics[width=0.9\linewidth]{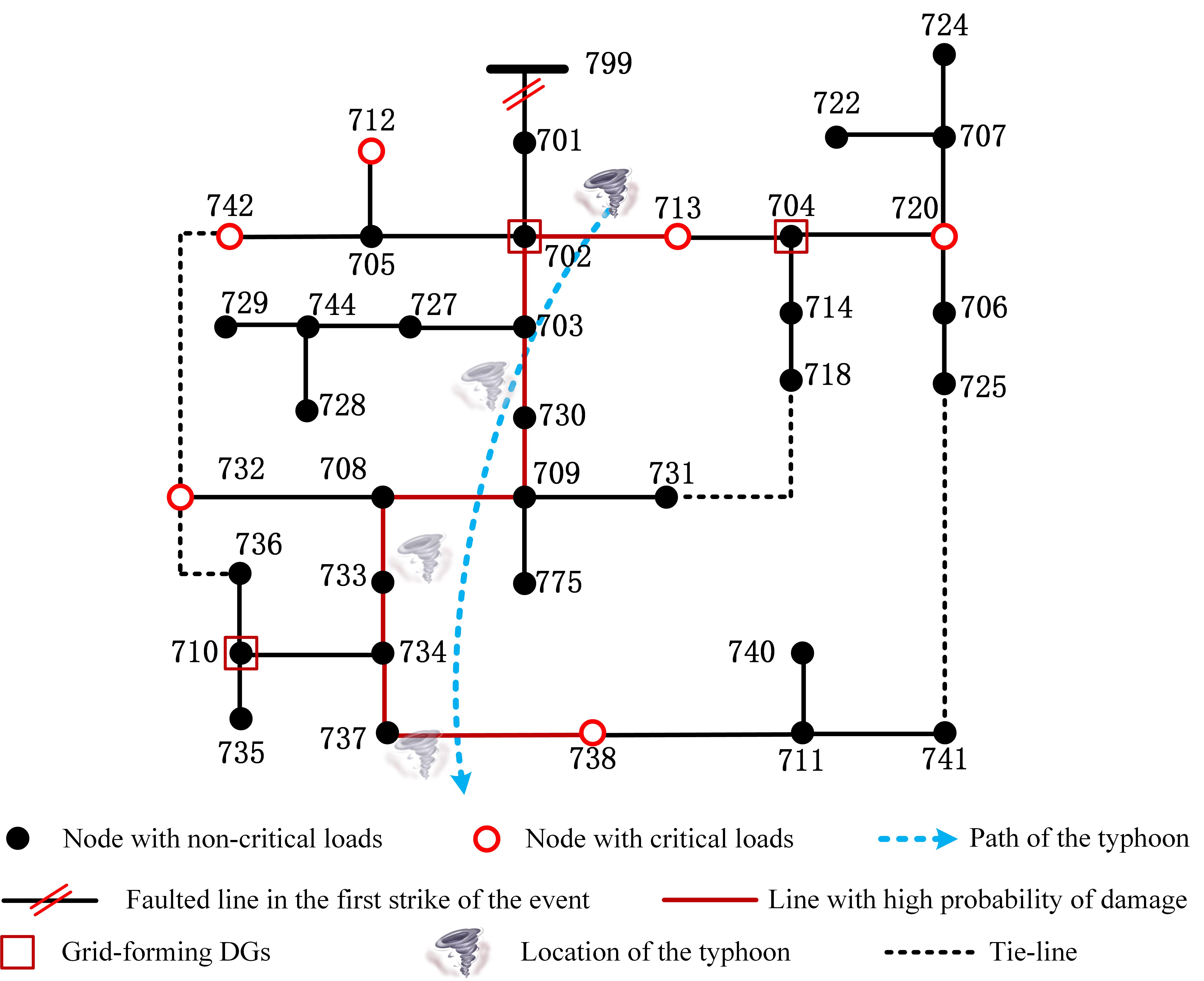}
    \caption{The modified IEEE 37-bus system}
    \label{fig_modified37}
\end{figure}
\section{Numerical Simulations}
In this section, the proposed DR-DMF model is demonstrated on a modified IEEE 37-node system\cite{cai2021distributionally}, which is shown in Fig. \ref{fig_modified37}.
Three grid-forming DGs are employed in the system, and applied to nodes 702, 704, and 710, respectively. 
Four tie-lines (736-742, 725-741, 732-736, and 718-731) are added into the system. 
The weight coefficient for critical loads and non-critical loads are set to \$100/kWh and \$10/kWh, respectively.
The trajectory of a typhoon is depicted in Fig. \ref{fig_modified37} with a blue arrow, while the lines along the trajectory are associated with a high probability of failure as the typhoon traverses through.
It is assumed that the substation sustains damage upon the initial impact of the typhoon, and the typhoon will persist for two hours. The time interval selected in the DR-DMF model is 30 minutes in this case, and the entire procedure is partitioned into four discrete time steps.

We use a computer with an Intel i7-1165G7 processor and 16GB memory. Simulations are implemented on Matlab with the commercial solver Gurobi 9.5.2.
\subsection{Optimization Results of the DR-DMF Model}
\begin{figure}[t]
    \centering
    \includegraphics[width=0.9\linewidth]{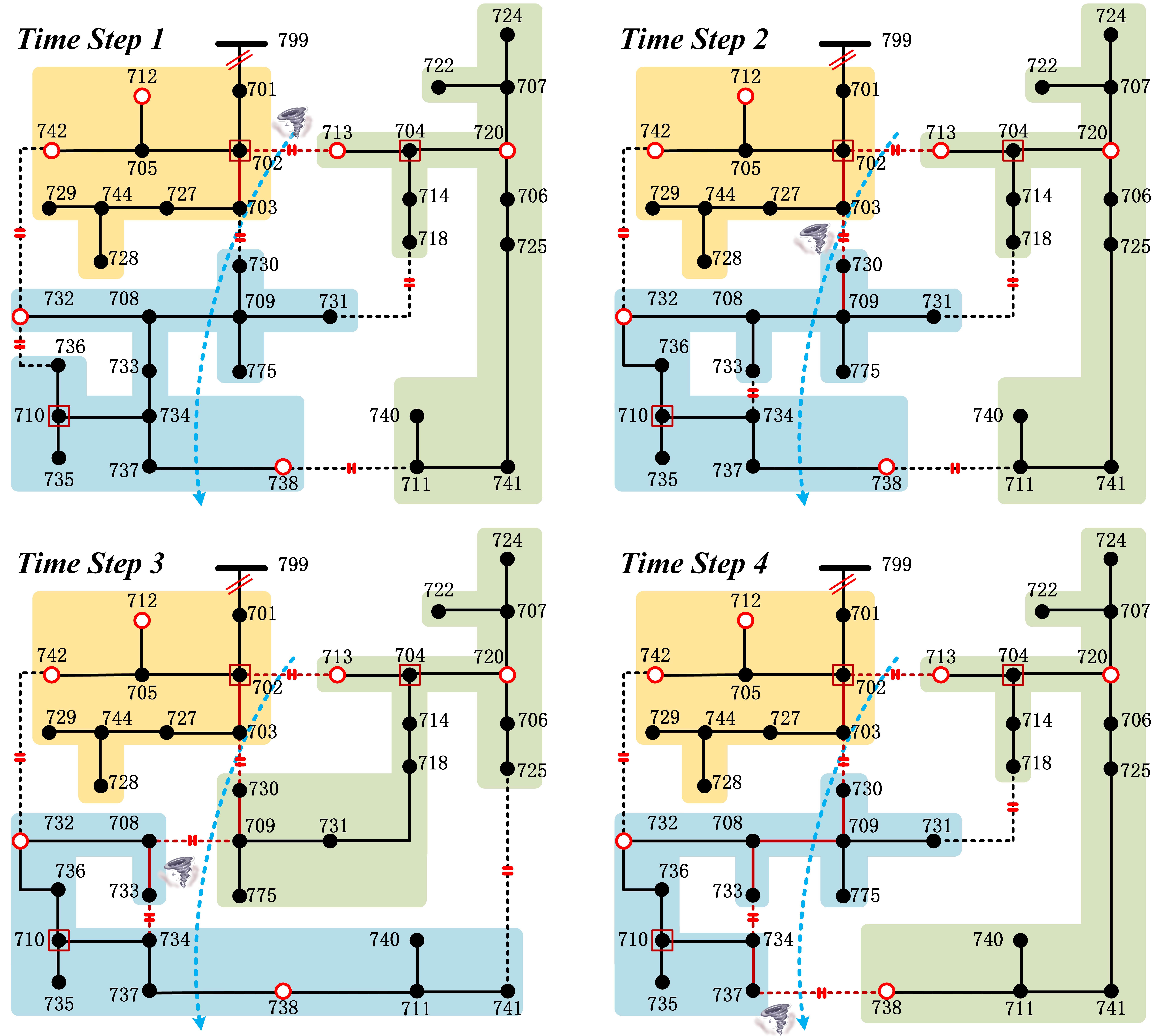}
\vspace{-1em}
    \caption{The result of the DR-DMF model in four time steps}
    \label{fig_result}
\vspace{-1em}
\end{figure}

Fig. \ref{fig_result} shows the DR-DMF result of four time steps. The distribution system is partitioned into three microgrids, due to the existence of three grid-forming DGs in the system. 
In the first time step, lines 702-713, 703-730, 711-738, 718-731, 732-736, and 732-742 are disconnected. 
The typhoon is traveling through line 702-713, which raises the failure probability of this line so that the line is disconnected. 
For the next three time steps, the boundaries of each microgrid are dynamically adjusted due to the trajectory of the typhoon. The lines with high failure probability will be probably excluded from each microgrid, which depends on the impact of the failure of the line. 
Moreover, all the critical loads are included in the microgrids, and the power supply path from each grid-forming DG to the critical loads is reliable. 
For instance, at the fourth time step, the critical load at node 738 is assigned to the green microgrid due to the high risk of line 737-738.

\subsection{Monte Carlo Simulation and Comparisons}
Notably, the microgrid formation decisions of the DR-DMF model are obtained prior to the occurrence of uncertainties. To assess the performance of the DR-DMF model, contingency scenarios must be generated. A small disturbance is added to the prior failure rate of each line to simulate the real distribution of line failure. Then, the Monte Carlo (MC) simulation is used to generate contingency scenarios. Since the line failure rate is relatively high in extreme weather, 1000 scenarios are generated based on the real distribution of the line failure contingencies. 

To further illustrate the performance of the proposed model, three cases are built and compared.
\begin{enumerate}
    \item Method \#1: the proposed DR-DMF model.
    \item Method \#2: the distributionally robust static microgrid formation (DR-SMF) model, where the boundaries of each microgrid are static.
    \item Method \#3: the robust dynamic microgrid formation model (\textit{N-k} criterion).
\end{enumerate}
\begin{figure}[t]
    \centering
    \includegraphics[width=0.9\linewidth]{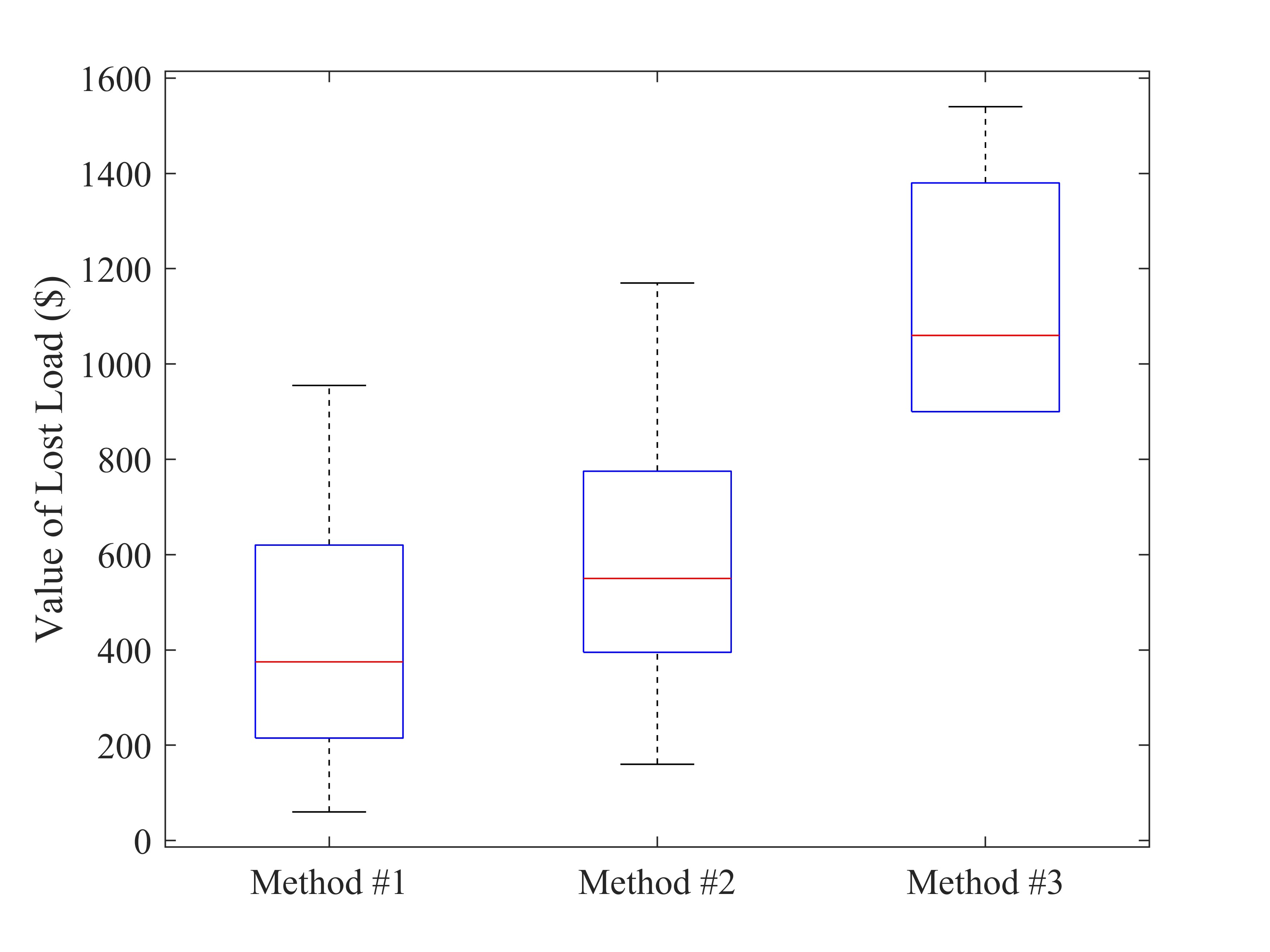}
    \vspace{-1em}
    \caption{Box-plot of weighted load shedding of three methods}
    \label{fig_boxplot}
\end{figure}
\begin{table}[t]
    \centering
    \caption{Expected Value of VoLL}
    \label{tab_E_load_sheding}
    \begin{tabular}{cccc}
        \hline
        \hline
        \textbf{Time Step} 
        &\multicolumn{1}{c}{\begin{tabular}[c]{@{}c@{}}\textbf{Method \#1}\\ \textbf{(\$)}\end{tabular}}
        &\multicolumn{1}{c}{\begin{tabular}[c]{@{}c@{}}\textbf{Method \#2}\\ \textbf{(\$)}\end{tabular}}
        &\multicolumn{1}{c}{\begin{tabular}[c]{@{}c@{}}\textbf{Method \#3}\\ \textbf{(\$)}\end{tabular}}
        \\
        \hline
        \textbf{1} & 58.9 & 218.9 & 778.9 \\
        \textbf{2} & 184.0 & 344.0 & 1130.9 \\
        \textbf{3} & 396.8 & 763.9 & 1246.1 \\
        \textbf{4} & 1053.2 & 1053.2 & 1337.6 \\
        \hline
        \textbf{Total} & 1692.9 & 2380.0 & 4493.5 \\
        \hline
        \hline
    \end{tabular}
\end{table}

Fig. \ref{fig_boxplot} depicts the box-plot graph of the value of lost load (VoLL) for three methods under all the scenarios. 
The VoLL of the three methods is shown in Table \ref{tab_E_load_sheding}. According to Table \ref{tab_E_load_sheding} and Fig. \ref{fig_boxplot}, it can be seen that: 

1) The proposed DR-DMF method performs much better than the other two methods. 
Five criteria of the box-plot show the superiority of the proposed model.
The total VoLL in four time steps is reduced by 28.9\% and 62.3\%, respectively. 

2) During the last time step, the simulation results of the DR-DMF method and Method \#2 are identical, due to the utilization of the same line failure rate information;
However, the information on time-dependent line failure rate is neglected in Method \#2 which leads to overly conservative results. 
Note that the conservative approach results in increased load shedding as it disregards the reliable power supply to non-critical loads.

3) The result of Method \#3 indicates that the RO model is much more conservative than the DRO model, even when compared to the DR-SMF model. It also demonstrates the effectiveness of adopting DRO in the DMF process.

\section{Conclusion}
This paper proposes a distributionally robust dynamic microgrid formation (DR-DMF) method. The temporal information of line failure probability is fully considered during long-duration extreme weather events, like typhoons. 
Furthermore, the expected weighted load shedding is minimized by the distributionally robust optimization model considering the uncertainty of line failure probability regarding the worst distribution of contingencies in each time interval. 
The proposed model is reformulated into a two-stage robust model and solved by the C\&CG method effectively.
Finally, the numerical simulations are tested based on a modified IEEE 37-bus system with tie-line switches. 
The simulation results demonstrate the effectiveness of the proposed DR-DMF method in substantially reducing the VoLL and enhancing system resilience under long-duration extreme weather events.
\bibliographystyle{ieeetr}
\bibliography{ref}

\begin{thebibliography}{10}

\bibitem{Texas_Snowstorm}
J.~Donald, ``Winter storm uri 2021, the economic impact of the storm.'' https://comptroller.texas.gov/economy/fiscal-notes/2021/oct/winter-storm-impact.php.

\bibitem{campbell2012weather}
R.~J. Campbell and S.~Lowry, ``Weather-related power outages and electric system resiliency,'' Congressional Research Service, Library of Congress Washington, DC, 2012.

\bibitem{lei2017remote}
S.~Lei, J.~Wang, and Y.~Hou, ``Remote-controlled switch allocation enabling prompt restoration of distribution systems,'' {\em IEEE Transactions on Power Systems}, vol.~33, no.~3, pp.~3129--3142, 2017.

\bibitem{shi2021network}
Q.~Shi, F.~Li, M.~Olama, J.~Dong, Y.~Xue, M.~Starke, C.~Winstead, and T.~Kuruganti, ``Network reconfiguration and distributed energy resource scheduling for improved distribution system resilience,'' {\em International Journal of Electrical Power \& Energy Systems}, vol.~124, p.~106355, 2021.

\bibitem{ding2017resilient}
T.~Ding, Y.~Lin, Z.~Bie, and C.~Chen, ``A resilient microgrid formation strategy for load restoration considering master-slave distributed generators and topology reconfiguration,'' {\em Applied energy}, vol.~199, pp.~205--216, 2017.

\bibitem{hussain2019microgrids}
A.~Hussain, V.-H. Bui, and H.-M. Kim, ``Microgrids as a resilience resource and strategies used by microgrids for enhancing resilience,'' {\em Applied energy}, vol.~240, pp.~56--72, 2019.

\bibitem{wang2015networked}
Z.~Wang, B.~Chen, J.~Wang, and C.~Chen, ``Networked microgrids for self-healing power systems,'' {\em IEEE Transactions on smart grid}, vol.~7, no.~1, pp.~310--319, 2015.

\bibitem{hussain2016resilient}
A.~Hussain, V.-H. Bui, and H.-M. Kim, ``A resilient and privacy-preserving energy management strategy for networked microgrids,'' {\em IEEE Transactions on Smart Grid}, vol.~9, no.~3, pp.~2127--2139, 2016.

\bibitem{kim2016framework}
Y.-J. Kim, J.~Wang, and X.~Lu, ``A framework for load service restoration using dynamic change in boundaries of advanced microgrids with synchronous-machine dgs,'' {\em IEEE Transactions on Smart Grid}, vol.~9, no.~4, pp.~3676--3690, 2016.

\bibitem{lei2020radiality}
S.~Lei, C.~Chen, Y.~Song, and Y.~Hou, ``Radiality constraints for resilient reconfiguration of distribution systems: Formulation and application to microgrid formation,'' {\em IEEE Transactions on Smart Grid}, vol.~11, no.~5, pp.~3944--3956, 2020.

\bibitem{cai2022hybrid}
S.~Cai, M.~Zhang, Y.~Xie, Q.~Wu, X.~Jin, and Z.~Xiang, ``Hybrid stochastic-robust service restoration for wind power penetrated distribution systems considering subsequent random contingencies,'' {\em IEEE Transactions on Smart Grid}, vol.~13, no.~4, pp.~2859--2872, 2022.

\bibitem{gholami2017proactive}
A.~Gholami, T.~Shekari, and S.~Grijalva, ``Proactive management of microgrids for resiliency enhancement: An adaptive robust approach,'' {\em IEEE Transactions on Sustainable Energy}, vol.~10, no.~1, pp.~470--480, 2017.

\bibitem{babaei2020distributionally}
S.~Babaei, R.~Jiang, and C.~Zhao, ``Distributionally robust distribution network configuration under random contingency,'' {\em IEEE Transactions on Power Systems}, vol.~35, no.~5, pp.~3332--3341, 2020.

\bibitem{cai2021distributionally}
S.~Cai, Y.~Xie, Q.~Wu, M.~Zhang, X.~Jin, and Z.~Xiang, ``Distributionally robust microgrid formation approach for service restoration under random contingency,'' {\em IEEE Transactions on Smart Grid}, vol.~12, no.~6, pp.~4926--4937, 2021.

\bibitem{lavorato2011imposing}
M.~Lavorato, J.~F. Franco, M.~J. Rider, and R.~Romero, ``Imposing radiality constraints in distribution system optimization problems,'' {\em IEEE Transactions on Power Systems}, vol.~27, no.~1, pp.~172--180, 2011.

\bibitem{jabr2012minimum}
R.~A. Jabr, R.~Singh, and B.~C. Pal, ``Minimum loss network reconfiguration using mixed-integer convex programming,'' {\em IEEE Transactions on Power systems}, vol.~27, no.~2, pp.~1106--1115, 2012.

\bibitem{zeng2013solving}
B.~Zeng and L.~Zhao, ``Solving two-stage robust optimization problems using a column-and-constraint generation method,'' {\em Operations Research Letters}, vol.~41, no.~5, pp.~457--461, 2013.

\bibitem{mccormick1976computability}
G.~P. McCormick, ``Computability of global solutions to factorable nonconvex programs: Part i—convex underestimating problems,'' {\em Mathematical programming}, vol.~10, no.~1, pp.~147--175, 1976.

\end{thebibliography}

\end{document}